# Regulation of Interfacial Chemistry by Coupled Reaction-Diffusion Processes in the Electrolyte: A Stiff Solution Dynamics Model for Corrosion and Passivity of Metals

Infant G. Bosco[a,b], Ivan S. Cole[a] and Bosco Emmanuel[c#]

[a] CSIRO Materials Science and Engineering, Clayton, 3169 Victoria, Australia

[b] Institute for Frontier Materials, Deakin University, Burwood, 3125 Victoria, Australia

[c] CSIR-CECRI, Modelling and Simulation, Karaikudi, 630006 Tamilnadu, India

**Abstract**

In this paper we advance a stiff solution dynamics [SSD] model to study the regulation of local chemistry near a corroding metal by reaction and diffusion processes in the electrolyte. Using this model we compute the detailed space-time dynamics of the concentrations of metal ions, its hydroxy complexes, $H^+$ and $OH^-$ ions near the corroding metal. The time for the onset of passivity for Fe and Zn is presented for free corrosion condition, different impressed currents and initial pH values. The theory advanced provides much physical insight into corrosion and passivity of metals and motivate spectro-electrochemical studies.

#corresponding author   e-mail: boscoemmanuel@yahoo.co.in

Phone +91-4565-227550   Fax   +91-4565-227779

# 1. Introduction

As early as 1972 Pickering and Frankenthal [1] modelled localised corrosion of iron and steel by considering the diffusion and migration of metal ions and hydrogen ions in artificial pits. Galvele and co-workers extended this model to include solution processes such as metal ion hydrolysis and self-hydrolysis of water and studied their role in passivity breakdown [2,3]. These are steady state models involving a known constant current due to metal dissolution at the bottom of the pit. While chemical reactions such as the metal ion hydrolysis leading to $H^+$ generation in the pit and self-hydrolysis of water are recognised, the cathodic reactions like oxygen reduction or hydrogen evolution and the consequent changes in the solution pH are not considered by these authors. Though these models have led to much useful insights into the conditions under which passivity sets in, two limitations of these models should be noted:

(1) These models describe only the steady state and consequently can not capture the time-dependent changes in the pit solution leading to eventual passivity or pitting. For this reason, they can not predict properties such as the time for passivity. Importantly passivity may set in before the steady state is reached.

(2) Only the anodic metal dissolution can be included in their models and the cathodic counter reactions like oxygen reduction:

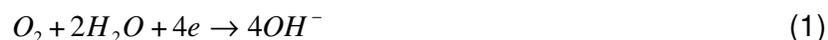
$$O_2 + 2H_2O + 4e \rightarrow 4OH^- \qquad (1)$$

cannot be included. The reason for the inability of these models to include any cathodic counter reaction can be traced to the fact that Galvele's model is based on the steady state "atom" fluxes and not on the "species" fluxes. For electrode reactions such as (1) where the reactant species as well as the product species are in the solution, the corresponding atom fluxes at the electrode surface turns out to be zero. For example, in the oxygen reduction reaction above, four hydrogen atoms and four oxygen atoms enter as part of the reactants ( $O_2$ and $2H_2O$ ) and the identical numbers leave as the product ( $4OH^-$ ). Though reaction (1) leads to a flux of $OH^-$ ions going into the solution, Galvele's model which is based on atom fluxes can not describe this flux. It can capture only the metal ion flux arising from the anodic dissolution of the metal:   $M \rightarrow M^{n+} + ne$ (2)

Here the metal M is a part of the electrode, while the metal ion is in the solution. Hence there is a non-zero metal atom flux at the metal/solution interface.

It is clear that for a complete understanding of the passivity phenomenon, the species fluxes (e.g. $OH^-$ and $H^+$) at the electrode surface arising from the cathodic counter reactions such as oxygen reduction and hydrogen evolution should also be included in the model besides the metal ion flux. The proposed SSD model is aimed at achieving this goal and provides a new theoretical methodology for describing the time-dependent changes in the solution composition leading to passivity or pitting. Unlike the earlier models which are applicable only to the impressed current condition the present model is applicable to both the free corrosion condition and the impressed current situation.

A typical corrosion scenario involves a metal or alloy surface generating a flux of metal ions and hydroxyl ions or consuming hydrogen ions.

The metal ions diffuse into the solution, hydrolyse water generating $H^+$ which in turn modify the self hydrolysis equilibrium of water ($H^+ + OH^- = H_2O$) and react with a host of other ions such as hydroxyl, chloride, bicarbonate and sulphate depending on the composition of the corrosive medium. When the concentration of hydroxy, chloro-hydroxy and other metal complexes exceed certain solubility limits passive layers may deposit on the metal. This will decide between passivity and pitting when these processes take place inside pits, cracks or other voids present in bare or coated metals. On uniform metallic surfaces, general corrosion or passivity will be the result. Precipitation and strong bonding of the precipitate to the corroding metal leading to a compact, non-porous layer will be ideal for corrosion control and self-repair. On the other hand, if the corrosion products are loosely adherent to the metal surface, soft and porous or if they precipitate in the solution, passivity will not set in. Therefore the question of if and when the solubility thresholds are exceeded become important. . In fact Cole et al. [4] undertook an experimental scanning electron microscope/ focussed ion beam and in-situ Raman study of oxide growth on zinc under seawater droplets. In the in-situ Raman they observed rapid (within minutes) growth in the Zn-O bond vibration and somewhat slower growth in sulphate and carbonate bond vibrations (probably associated with gordiate and hydrozincite). The focused ion beam sections demonstrated that solid solution growth of the oxide initially dominated with the growth of a high porous crystalline phase (gordiate or simonkolleite) or precipitation of crystalline phase from solution occurring after some time ( around 30 minutes) . The present model is aimed at capturing, in such situations, the solution dynamics leading to passivity or pitting.

In the present work we consider two different geometries: semi-infinite and finite. For corrosion in bulk electrolytes the semi-infinite geometry will be appropriate while the finite geometry will be useful for a metal covered with a thin electrolyte layer [5] or a porous oxide layer [6] and for the Rotating Disc Electrode [7].However detailed results are presented only for the semi-infinite geometry and work on the other two geometries is in progress [8].

In section 2, we formulate the detailed mathematical model with its assumptions and approximations clearly spelt out and the analytical solutions of the model for the space-time dependence of the various species concentrations are presented in section 3. Results, based on this model, for the time evolution of the surface concentration of the metal-ion complexes which can deposit and passivate the metal are presented and discussed in section 4 for free corrosion condition, different impressed currents and initial pH values for iron and zinc. Typical concentration-versus-distance profiles are also provided for all the species. Conclusions and future perspectives are in section 5.

**2. The SSD Model Framework and the Assumptions**

The model starts with known fluxes of metal ions and hydroxyl or hydrogen ions at the metal/electrolyte interface. For free corrosion condition the current densities associated with these fluxes balance one another so that there is no net current through the system whereas for the case of impressed current/potential these fluxes will be such as to produce a net current through the system. We report on both these cases. Without loss of generality we treat here the case where oxygen reduction is the cathodic reaction while the case of hydrogen evolution will be taken up in the future work. Thus we have metal ions and hydroxyl ions coming into the solution where they diffuse and undergo solution reactions which are, in the simplest case, the following hydrolysis reactions involving the metal ion M²⁺ and the self-hydrolysis reaction of water.

$$M^{2+} + H_2O \Leftrightarrow (MOH)^+ + H^+ \tag{3}$$

$$(MOH)^+ + H_2O \Leftrightarrow M(OH)_2 + H^+ \tag{4}$$

$$H_2O \Leftrightarrow H^+ + OH^- \tag{5}$$

$M^{2+}$ may be any divalent metal ion such as $Fe^{2+}$, $Zn^{2+}$ and $Mg^{2+}$. For the trivalent metal ions $Al^{3+}$ and $Fe^{3+}$, there will be one more hydrolysis step. After labelling the species $M^{2+}$, $H_2O$, $(MOH)^+$, $H^+$, $M(OH)_2$ and $OH^-$ respectively by the numerals 1,2,3,4,5 and 6, the stability constants may be written as

$$K_1 = \frac{C_3 C_4}{C_1 C_2} \tag{6}$$

$$K_2 = \frac{C_4 C_5}{C_2 C_3} \tag{7}$$

$$K_3 = \frac{C_4 C_6}{C_2} \tag{8}$$

Where $C_i$ is the concentration of the $i^{th}$ species in the reactions (3) to (5) above.

The species 1 to 6 diffuse in the solution and while diffusing they also undergo chemical reactions. In addition, some of these species will be generated or consumed at the electrode surface by corrosion reactions and their influence on the chemical reactions in the electrolyte will enter through appropriate boundary conditions for the reaction-diffusion equations for the species 1-6:

$$\frac{\partial C_1}{\partial t} = D_1 \frac{\partial^2 C_1}{\partial x^2} - R_1 \tag{9}$$

$$\frac{\partial C_2}{\partial t} = D_2 \frac{\partial^2 C_2}{\partial x^2} - (R_1 + R_2 + R_3) \tag{10}$$

$$\frac{\partial C_3}{\partial t} = D_3 \frac{\partial^2 C_3}{\partial x^2} + R_1 - R_2 \tag{11}$$

$$\frac{\partial C_4}{\partial t} = D_4 \frac{\partial^2 C_4}{\partial x^2} + (R_1 + R_2 + R_3) \tag{12}$$

$$\frac{\partial C_5}{\partial t} = D_5 \frac{\partial^2 C_5}{\partial x^2} + R_2 \tag{13}$$

$$\frac{\partial C_6}{\partial t} = D_6 \frac{\partial^2 C_6}{\partial x^2} + R_3 \tag{14}$$

Where $R_1$, $R_2$ and $R_3$ are the three chemical reaction rates corresponding to the reactions (3) to (5):

$$R_1 = k_{1f} C_1 C_2 - k_{1b} C_3 C_4 \tag{15}$$

$$R_2 = k_{2f} C_2 C_3 - k_{2b} C_4 C_5 \tag{16}$$

$$R_3 = k_{3f} C_2 - k_{3b} C_4 C_6 \tag{17}$$

Now we make the important assumption that the diffusion coefficients of the species are the same. This approximation will be good if the species have nearly equal diffusion coefficients and if not it will provide lower and upper bounds for the space and time dependent concentrations of the species by choosing the lowest or the highest of the diffusion coefficients as the common diffusion coefficient. This approximation is well known in Electrochemistry. Besides Electrochemistry it has been used by several investigators in the field of bioengineering. These workers have argued that a diffusion potential gradient [9] develops which enhances the transport of the faster ions and this reduces the differences in the diffusivities of the individual species. This approximation is further justified by the fact that the transport due to reaction-diffusion coupling is much more significant that caused by the differences in the individual diffusivities. Let us denote the common diffusion by D.

The second assumption we make is that the chemical reactions take place on much faster time scales than the diffusion of species and hence we propose to apply the equilibrium constraints (6)-(8) for the species concentrations. This assumption can be justified on the basis that hydrolysis reactions are very rapid [10] and equilibrium will be achieved within microseconds in comparison with the diffusion process which takes typically 10 seconds to have its influence over a distance of 100 micrometers. [Besides we neglect poly-nuclear species as their concentrations will be low compared to the mononuclear species and their formation is a much slower process [11]]. Indeed what are known about these reactions are their equilibrium constants and not the individual forward and backward rate constants. Such problems with widely varying time scales are termed "stiff" in the mathematical literature and hence the name "Stiff Solution Dynamics" for the present model. This stiffness implies that, at every space and time point, we may equilibrate the species by subjecting the species concentrations to the equilibrium constraints (6)-(8) before the concentrations begin to change due to diffusion.

However, when we apply this equilibrium, we must ensure that the concentration of each atom (constituting the species) at every space-time point is the same before and after the equilibration step, as every chemical reaction is only a rearrangement of atoms among the reactant and product species which conserve the number of atoms of each kind. Now for the species concentration $C_1(x,t)$, $C_2(x,t)$, $C_3(x,t)$, $C_4(x,t)$, $C_5(x,t)$ and $C_6(x,t)$, the corresponding atom concentrations for the atoms of $M$, $H$ and $O$ are given by the following linear combinations. These linear combinations need to include only those species involved in the solution reactions. The dissolved $O_2$ concentration in the oxygen reduction reaction or the dissolved $H_2$ concentration in the hydrogen evolution reaction, even if included, does not alter the final result in a self-consistent calculation.

$$L_M(x,t) = C_1(x,t) + C_3(x,t) + C_5(x,t) \tag{18}$$

$$L_H(x,t) = 2C_2(x,t) + C_3(x,t) + C_4(x,t) + 2C_5(x,t) + C_6(x,t) \tag{19}$$

$$L_O(x,t) = C_2(x,t) + C_3(x,t) + 2C_5(x,t) + C_6(x,t) \tag{20}$$

We now come to an interesting stage in our development: if we use the above linear combinations in the reaction-diffusion equations (9) to (14), we find that the three linear combinations $L_M(x,t)$, $L_H(x,t)$ and $L_O(x,t)$ satisfy the following **diffusion-only** equations where the reaction terms have completely disappeared.

$$\frac{\partial L_M(x,t)}{\partial t} = D \frac{\partial^2 L_M(x,t)}{\partial x^2} \tag{21}$$

$$\frac{\partial L_H(x,t)}{\partial t} = D \frac{\partial^2 L_H(x,t)}{\partial x^2} \tag{22}$$

$$\frac{\partial L_O(x,t)}{\partial t} = D\frac{\partial^2 L_O(x,t)}{\partial x^2} \tag{23}$$

Though the above three partial differential equations (PDE's) are identical, their initial and boundary conditions will all be different. Therefore the solutions $L_M(x,t)$, $L_H(x,t)$ and $L_O(x,t)$ will be quite different from one another. The initial condition will be determined by the initial composition of the electrolyte whereas the boundary conditions on the metal depend on the corrosion reactions taking place and the participation or non- participation of the species.

It is to be emphasised that the solution of the PDE's (21) to (23) for the linear combinations (of species concentrations) $L_M(x,t)$, $L_H(x,t)$ and $L_O(x,t)$ are exact even in the presence of chemical reactions (3) to (5). However, these linear combinations do not as yet possess any information about the chemical reactions and apply irrespective of whether these chemical reactions are in equilibrium or kinetically driven. Once chemical equilibrium is assumed, we have the following six equations for the six species concentrations:

$$L_M(x,t) = C_1(x,t) + C_3(x,t) + C_5(x,t) \tag{24}$$

$$L_H(x,t) = 2C_2(x,t) + C_3(x,t) + C_4(x,t) + 2C_5(x,t) + C_6(x,t) \tag{25}$$

$$L_O(x,t) = C_2(x,t) + C_3(x,t) + 2C_5(x,t) + C_6(x,t) \tag{26}$$

$$K_1 = \frac{C_3(x,t)C_4(x,t)}{C_1(x,t)C_2(x,t)} \tag{27}$$

$$K_2 = \frac{C_4(x,t)C_5(x,t)}{C_2(x,t)C_3(x,t)} \tag{28}$$

$$K_3 = \frac{C_4(x,t)C_6(x,t)}{C_2(x,t)} \tag{29}$$

The RHS of equations (24) - (26) are known by solving the set of the three PDE's (21) - (23). $K_1$, $K_2$ and $K_3$ are known equilibrium constants. It is to be noted that the time and space dependence of the species concentrations arise from the time and space dependence of the linear combinations $L_M(x,t)$, $L_H(x,t)$ and $L_O(x,t)$.

## 3. Analytic solutions for the reaction diffusion model

To solve the PDE'S (21) to (23), we need to prescribe the initial and boundary conditions on $L_M(x,t)$, $L_H(x,t)$ and $L_O(x,t)$.

Initial Conditions:

$$L_M(x,0) = L_{M,0} \tag{30}$$

$$L_H(x,0) = L_{H,0} \tag{31}$$

$$L_O(x,0) = L_{O,0} \tag{32}$$

Where $L_{M,0}$, $L_{H,0}$ and $L_{O,0}$ may be found from the initial concentrations of the solution species using the equations (24) to (26).

Boundary Condition at the electrode surface at $x=0$:

$$-D\frac{\partial L_M}{\partial x} = F_{M^{2+}} = F_1 \tag{33}$$

$$-D\frac{\partial L_H}{\partial x} = 0 \tag{34}$$

$$-D\frac{\partial L_O}{\partial x} = \frac{F_{OH^-}}{2} = \frac{F_6}{2} \tag{35}$$

These conditions follow from the equations (24) to (26) and the stoichiometry of the oxygen reduction reaction:

$$O_2 + 2H_2O + 4e \rightarrow 4OH^-$$

Further for free corrosion condition with zero net current, $F_1 = \dfrac{F_6}{2}$. For an impressed current experiment $F_{M^{2+}}$, $F_{OH^-}$ and the impressed current density $I$ are related by

$$2F\ F_{M^{2+}} - F\ F_{OH^-} = I \qquad (36)$$

Which means that $F_{OH^-}$ may be known from $I$ and the metal ion flux $F_{M^{2+}}$. Here $F$ is the Faraday.

For boundary conditions at $x = \infty$, we should consider the specific experimental situation at hand. We consider 3 possible situations in this paper: (I) Diffusion and Reaction in the semi-infinite medium, (II) Diffusion and Reaction in a finite diffusion layer as in the Rotating Disc Electrode (RDE) and (III) Diffusion and Reaction in a thin electrolyte layer on the corroding metal. The boundary conditions and hence the solutions are different in each one of these cases. We outline the method of solution for Case (I) below and state the modifications necessary for Cases (II) and (III) towards the end of this section. Detailed results are presented below for Fe and Zn for Case(I) only and the results for Case(II) and Case (III) will be reported elsewhere [8].

Case (I): for diffusion and reactions in the semi-infinite medium the boundary conditions at $x = \infty$ follow easily from the initial conditions:

$$L_M(\infty, t) = L_{M,0} \qquad (37)$$

$$L_H(\infty, t) = L_{H,0} \qquad (38)$$

$$L_O(\infty,t) = L_{O,0} \qquad (39)$$

Using the method of Laplace transformation, the solutions to equations (21) to (23) can easily be obtained and they are:

$$L_M(x,t) = L_{M,0} + F_1 \left( 2\sqrt{\frac{t}{\pi D}} \exp\left(-\frac{x^2}{4Dt}\right) - \frac{x}{D} \text{erfc}\left(\frac{x}{2\sqrt{Dt}}\right) \right) \qquad (40)$$

$$L_H(x,t) = L_{H,0} \qquad (41)$$

$$L_O(x,t) = L_{O,0} + \frac{F_6}{2} \left( 2\sqrt{\frac{t}{\pi D}} \exp\left(-\frac{x^2}{4Dt}\right) - \frac{x}{D} \text{erfc}\left(\frac{x}{2\sqrt{Dt}}\right) \right) \qquad (42)$$

Now the RHS of equations (24) to (26) are known for every space-time point and hence the 6 equations (24) to (29) may be solved for the 6 species concentrations with their time and space dependences included through equations (40) to (42). We sketch the method of solution below:

Solving equations (24), (27) and (28) for $C_1(x,t)$, $C_3(x,t)$ and $C_5(x,t)$ we obtain:

$$C_1(x,t) = \frac{L_M(x,t)}{(1+A(x,t)+A(x,t)B(x,t))} \qquad (43)$$

$$C_3(x,t) = \frac{L_M(x,t)A(x,t)}{(1+A(x,t)+A(x,t)B(x,t))} \qquad (44)$$

$$C_5(x,t) = \frac{L_M(x,t)A(x,t)B(x,t)}{(1+A(x,t)+A(x,t)B(x,t))} \qquad (45)$$

Where

$$A(x,t) = \frac{K_1 C_2(x,t)}{C_4(x,t)} = \frac{K_1 C_6(x,t)}{K_3} = pC_6(x,t) \qquad (46)$$

and

$$B(x,t) = \frac{K_2 C_2(x,t)}{C_4(x,t)} = \frac{K_2 C_6(x,t)}{K_3} = qC_6(x,t) \qquad (47)$$

Thus we have expressed $C_1(x,t)$, $C_3(x,t)$ and $C_5(x,t)$ in terms of $C_6(x,t)$.

Subtraction of equation (26) from equation (25) results in

$$C_2(x,t) + C_4(x,t) = L_H - L_O \qquad (48)$$

Use equation (29) in equation (48) to obtain

$$C_2(x,t) = \frac{C_6(x,t)(L_H - L_O)}{(K_3 + C_6(x,t))} \qquad (49)$$

Substitute this in equation (26), $C_3(x,t)$ and $C_5(x,t)$ from equations (44) and (45) to obtain:

$$a_4 C_6^4 + a_3 C_6^3 + a_2 C_6^2 + a_1 C_6 + a_0 = 0 \qquad (50)$$

Where

$$a_0 = -K_3 L_O \qquad (51)$$

$$a_1 = L_H - 2L_O + K_3(1 + pL_M - pL_O) \qquad (52)$$

$$a_2 = 1 + p(L_H - 2L_O + L_M + 2qK_3 L_M - qK_3 L_O + K_3) \qquad (53)$$

$$a_3 = pq(L_H - 2L_O + 2L_M + K_3 + \frac{1}{q}) \qquad (54)$$

$$a_4 = pq \qquad (55)$$

Equation (50) is a quartic equation. Though in principle we can track the positive real root of this quartic equation, it will be quite messy and hence we have taken the simpler alternative of numerically computing the solution of equation (50) by using the "fsolve" command in MAPLE.

For the experimental Cases (II) and (III) all the preceding mathematical steps remain the same except the expressions for $L_M(x,t), L_H(x,t)$ and $L_O(x,t)$ which are modified as under.

Case (II):

For the RDE experiment, the following boundary conditions hold at $x = l$ where $l$ is the thickness of the diffusion layer that can be controlled by the speed of rotation of the RDE.

$$L_M(l,t) = L_{M,0} \tag{56}$$

$$L_H(l,t) = L_{H,0} \tag{57}$$

$$L_O(l,t) = L_{O,0} \tag{58}$$

For this Case and Case (III), the solutions can be found from one dimensional Greens functions [12] and some identities for trigonometric series [13]. The solutions for Case (II) are:

$$L_M(x,t) = L_{M,0} + \frac{F_1(l-x)}{D} - \frac{8F_1 l}{\pi^2 D} \sum_{n=0}^{\infty} \frac{\cos\left(\frac{(2n+1)\pi x}{2l}\right)\exp\left(-\frac{\pi^2 Dt(2n+1)^2}{4l^2}\right)}{(2n+1)^2} \tag{59}$$

$$L_H(x,t) = L_{H,0} \tag{60}$$

$$L_O(x,t) = L_{O,0} + \frac{F_6(l-x)}{2D} - \frac{4F_6 l}{\pi^2 D} \sum_{n=0}^{\infty} \frac{\cos\left(\frac{(2n+1)\pi x}{2l}\right)\exp\left(-\frac{\pi^2 Dt(2n+1)^2}{4l^2}\right)}{(2n+1)^2} \tag{61}$$

For the thin layer experiment, Case (III), the following boundary conditions hold at $x = l$ where $l$ is the thickness of the thin electrolyte layer on the corroding metal.

$$\frac{\partial L_M(x,t)}{\partial x} = 0 \tag{62}$$

$$\frac{\partial L_H(x,t)}{\partial x} = 0 \tag{63}$$

$$\frac{\partial L_O(x,t)}{\partial x} = 0 \tag{64}$$

The solutions for Case (III) are:

$$L_M(x,t) = L_{M,0} + \frac{F_1 t}{l} + \frac{F_1 l}{D}\left(\frac{x^2}{2l^2} - \frac{x}{l} + \frac{1}{3}\right) - \frac{2F_1 l}{\pi^2 D}\sum_{m=1}^{\infty}\frac{\cos\left(\frac{m\pi x}{l}\right)\exp\left(-\frac{\pi^2 m^2 D t}{l^2}\right)}{m^2} \tag{65}$$

$$L_H(x,t) = L_{H,0} \tag{66}$$

$$L_O(x,t) = L_{O,0} + \frac{F_6 t}{2l} + \frac{F_6 l}{2D}\left(\frac{x^2}{2l^2} - \frac{x}{l} + \frac{1}{3}\right) - \frac{F_6 l}{\pi^2 D}\sum_{m=1}^{\infty}\frac{\cos\left(\frac{m\pi x}{l}\right)\exp\left(-\frac{\pi^2 m^2 D t}{l^2}\right)}{m^2} \tag{67}$$

This completes the solution process.

## 4. Results and Discussions

The framework we developed here does not yet take account of migration in an electrical field. Hence the model, in its present state of development, is applicable to two experimental situations: (A) the free corrosion condition where there is no current or potential gradients (except of course in the double layer region) in the solution and (B) the impressed current or potential experiment where the solution has enough supporting electrolyte and only diffusion of species under consideration is important. In the concluding section, we briefly point out how the present model can easily incorporate convection besides diffusion and reaction.

In this paper we report the results for iron and zinc. The theory and computations for Mg shall proceed along similar lines whereas Al will need one more hydrolysis step as it generates trivalent Al ions in the solution. Though we are in this paper concerned with the initial precipitation of ferrous hydroxide this will be eventually oxidised to ferric oxide or magnetite or lose a water molecule to form FeO, the ferrous oxide. Another possibility which we are not presently considering is the formation ferric hydroxide which may lead to ferric oxide upon removal of two water molecules. This possibility which again needs the third hydrolysis step is to be taken up in later work. For the purposes of this paper we need the equilibrium constants of the first and second hydrolysis steps $K_1$ and $K_2$. $K_1$ is reported by Sillen et al [14] and Gravano and Galvele [3]. As we need $K_2$ also, we used the cumulative or gross constants $\beta_1$ and $\beta_2$ reported by Sillen et al [14] for the reaction of hydroxyl ions with the metal ions and computed $K_1$ and $K_2$. The constants calculated thus are $1.8 \times 10^{-11}$ and $1.8 \times 10^{-9.9}$ for the zinc system and $10^{-5.92}$ and $18^2 \times 10^{-18.91}$ for the iron system. The equilibrium constant for the self-hydrolysis of water is $1.8 \times 10^{-16}$ mole / (dm)$^3$. The solubility products of ferrous hydroxide and zinc hydroxide are respectively $4.87 \times 10^{-17}$ and $4.5 \times 10^{-17}$ in the appropriate units.

Figures 1(a) to 1(h) shows the time dependence of the concentrations of all the species near a corroding Iron surface and some relevant concentration products [Note that the concentration unit used throughout this paper is mole/dm³]. The concentration of $Fe^{2+}$ rises from zero sharply and becomes steady within seconds while that of $Fe(OH)^+$, $Fe(OH)_2$ and $OH^-$ rise gradually. The incoming hydroxyl ions are captured by the $H^+$ ions in the electrolyte with a sudden fall in the $H^+$ concentration. The ionic product of water remains independent of time as it should in figure 1(f). The horizontal line in figure 1(g) is the solubility product of ferrous hydroxide whose intersection with the concentration product $Fe^{2+}xOH^-xOH^-$ gives the time for the onset of oxide formation. This oxide formation is a precondition for passivity. The extent of passivity will be inferred by the nature of the oxide. Throughout the paper, the time for the onset of oxide formation will be referred to a passivation time. The concentration of water in figure 1(h) decreases with time and the change is in the fourth decimal as expected.

Figures 2(a) to 2(c) display the dependence of the time of onset of passivation (hereafter referred to as passivation time) on the corrosion current density when the latter varies from $2.08 \times 10^{-5}$ A/(dm)², $2.08 \times 10^{-4}$ A/(dm)² to $2.08 \times 10^{-3}$ A/(dm)² and the corresponding passivation times are 25 days, 6 hours and 4 minutes. Note that $2.08 \times 10^{-4}$ A/(dm)² is a typical value for iron. The time of passivity is indeed very sensitive to the corrosion current. Though the corrosion current varies from metal to metal the case we discuss here is the possible variation for the same metal arising from different experimental conditions.

Figure 3 has the dependence of passivation time for different impressed currents ranging from $-1.04 \times 10^{-4}$ A/(dm)² to $+1.04 \times 10^{-4}$ A/(dm)². Note that under the free corrosion condition the passivation time is about 6 hours while for cathodic polarisation it is lower at 3 and 4 hours.

Anodic polarisation leads to passivation for one case while the system does not passivate at all for the other case in the time window of the plot in figure 3.

Using the present SSD model concentration versus distance plots were also generated which provide an idea of how the concentrations of species and their products vary as we go away from the corroding metal. Figures 4(a) to 4(f) show this variation at time = 6.9 hours. The metal ion concentration decreases in a step-like fashion as we move away from the metal surface. The concentrations of $Fe(OH)^+$, $Fe(OH)_2$ and water vary monotonically to reach their bulk values. The product $Fe^{2+}xOH^-xOH^-$ decreases monotonically starting from a value above the solubility limit. While figure 4(e) corresponds to a time of 6.9 hours, figure 4(f) gives the variation at 6.1 hours when the system has not yet crossed the solubility limit.

Now the results for zinc: The trends are qualitatively the same as that for iron except that the passivation times are much shorter varying from 10 milliseconds to 100 seconds as the corrosion current varies from 4.58x10^(-3) to 4.58x10^(-5) A/(dm)$^2$ with 4.58x10^(-4) A/(dm)$^2$ being typical for zinc. Figures 5(a) to 5(c) has the concentration-product-versus-time plots for zinc. As mentioned in the introduction, the study of Cole et al [4] showed that oxide growth occurred within minutes of placing a saline droplet on a zinc surface. Although the geometry is different (drop as opposed to bulk electrolyte) the rapid growth observed is consistent with the predictions in this paper. Experimental work on iron is in progress.

The pH dependence was also probed. It is known that the corrosion rate, which enters the present model as the corrosion current/the metal-ion flux, is independent of pH(for the iron system) in the pH range 4 to 10 whereas it is controlled by hydrogen evolution below pH 4. As we are presently considering metal dissolution and oxygen reduction, the range of pH 4 to 10 will only be appropriate. The case of hydrogen evolution is also interesting in that the ionic fluxes at the corroding metal surface will change with the interfacial pH and hence time-dependent in contrast to the case of oxygen reduction. This case will be taken up later. The pH dependence of the time of passivation for Iron is as follows: 7.22 hours for pH = 4, nearly 6.25 hours for pH: 5 to 9 and 5 hours for pH = 10. To conserve space the detailed figures are omitted.

An assumption we have made to make the model solvable is that the diffusing species have a common diffusion coefficient D. Without this assumption the reaction-diffusion model can not be solved as the reaction rates of the metal-ion hydrolysis and the self-hydrolysis of water are not known. Though we have used the typical diffusion coefficient value of $10^{-5}$, we would recommend treating the effective value of this common diffusion coefficient as an experimentally adjustable parameter. As the passivation time is sensitive to the value of this effective diffusion coefficient, the kinetic plots generated in the present work must be understood as providing the qualitative trends while the quantification must come from correlating the experimentally observed time scales of passivation with this effective diffusion coefficient. Further as was pointed out earlier the present SSD model is applicable to free corrosion condition for dilute and concentrated electrolytes while the impressed potential/current condition would make the model inapplicable to dilute electrolytes where migration is important. However the model will hold exactly for convective transport as the convective flux is the product of the fluid velocity and the concentration and any linear combination of concentrations will be preserved in the flux and in the transport equations. Consequently the present model will find a good use in problems involving chemical reactions and fluid flow as in effluent discharges. Following earlier workers [2, 3] we have assumed that the metal-ion hydrolysis and self-hydrolysis reactions are in chemical equilibrium which we believe is a good approximation. Besides we can not presently consider the full chemical kinetics as we do not know the corresponding reaction rates. Nonetheless, unlike the steady state model of Galvele and co-workers, the present model can capture even those experimental situations where passivity may set in before the steady state is attained.

In this work we have treated dissolution-diffusion-reaction-and-passivity involving divalent metal ions such as $Fe^{2+}$ and $Zn^{2+}$ with the oxygen reduction as the counter reaction. The generalisations of the model involving trivalent metal ions like $Fe^{3+}$ and $Al^{3+}$ with hydrogen evolution as the possible counter reaction are in progress [8]. Care must be exercised in the proper choice of the counter reaction. For example, in the case of iron, the counter reaction is oxygen reduction between pH 4 and pH 10 and it is hydrogen evolution below a pH of 4. In addition, the corrosion rate of iron linearly decreases with pH below a pH of 4 and our model can handle this variation in the corrosion rate with a time-dependent pH near the metal [8].

For the sake of simplicity and clarity of presentation only a small number of species and reactions were considered in the electrolyte. For electrolytes containing other anions like chloride, bicarbonate and sulphate additional reactions will also be relevant. In the presence of chloride ions, for example, chloro- and chloro-hydroxy complexes of the metal need to be included in the reaction scheme and these complexes, when they are more soluble than the simple hydroxy complexes, could even delay the onset of passivity which in turn could be used to explain the role of chloride ion in corrosion. The necessary generalisations to incorporate these other anions and reactions involving them are available with us [8] and they will be reported later.

**5. Conclusions and Future Perspectives**

The present work provides a strong motivation for an experimental program that will probe the space and time dependent species concentrations in the solution near corroding metal and alloy surfaces using spectro-electrochemical methods [15, 16]. This will greatly compliment studies such as fib-sem for the solid corrosion products presently undertaken by corrosion researchers. To sum-up the present work:

- In this paper we advance a stiff solution dynamics [SSD] model to study the regulation of local chemistry near a corroding metal by reaction and diffusion processes in the electrolyte.
- Using this model we compute the detailed space-time dynamics of the concentrations of metal ions, its hydroxy complexes, $H^+$ and $OH^-$ ions.
- Map the detailed time evolution of the chemistry near the corroding metal generating metal ion and hydroxyl fluxes.
- The time for the onset of passivity for Fe and Zn is presented for free corrosion condition, different impressed currents and initial pH values.

- On Zn passivity sets in at much shorter time scales compared to Fe.
- Unlike the steady state model of Galvele and co-workers, the present dynamic model can capture even those experimental situations where passivity sets in before the steady state is attained.
- The new theoretical methodology advanced in this paper provides much physical insight into corrosion and passivity of metals.

Though the model developed here is applied to corrosion, its utility is more general. The more general setting in which this model will be useful consists of any interface (chemical, electrochemical and biological in nature) generating a flux of charged or neutral species which go into the immediate environment, diffuse and react with other species present. Electrocatalysis, Enzymatic Kinetics, Batteries and Fuel Cells, and Industrial/ Environmental Chemistry are a few areas, besides corrosion, where the proposed model will be relevant. In particular, pH generation and control [17] by diffusion and reaction is crucial in many systems including the human body.

**References**

[1] H.W. Pickering, R.P. Frankenthal, On the mechanism of localized corrosion of iron and stainless steel, J.Electrochem.Soc. 119(1972) 1297-1304.

[2] J.R. Galvele, Transport processes and the mechanism of pitting of metals, J.Electrochem.Soc.123(1976) 464-474.

[3] S.M. Gravano, J.R. Galvele, Transport processes in passivity breakdown III: full hydrolysis plus ion migration plus buffers, Corrosion Science, 24 (1984)517-534.

[4] I.S. Cole, T.H. Muster, Products formed during the interaction of sea-water droplets with zinc surfaces, J.Electrochem.Soc. 157(2010) C213-C222.

[5] M. Venkatraman, I.S. Cole, B. Emmanuel, Model for corrosion of metals covered with thin electrolyte layers: Pseudo-steady state diffusion of oxygen, Electrochim.Acta. 56(2011)7171-7179.

[6] M. Venkatraman, I.S. Cole, B. Emmanuel, Corrosion under a porous layer: A porous electrode model and its implications for self-repair, Electrochim.Acta. 56(2011)8192-8203.

[7] V.G. Levich, Physicochemical Hydrodynamics, Prentice Hall, New Jersey, USA, 1962.

[8] B. Emmanuel, work in progress.

[9] E. Ruckenstein, S. Varanasi, Acid generating immobilized enzymatic reactions in porous media : Activity control via augmentation of proton diffusion by weak acids, Chem.Eng.Sci. 39(1984)1185-1200.

[10] C.F. Baes, R.E. Mesmer, The Hydrolysis of Cations, John Wiley, New York, USA, 1976.

[11] J.P. Hunt, Metal Ions in Aqueous Solution, Benjamin, New York, USA, 1963.

[12] J.V. Beck, K.D.Cole, A. Haji-Sheikh, B. Litkouhi, Heat Conduction Using Green's Functions, Taylor & Francis, New York, USA, 1992.

[13] I.S. Gradshteyn, I.M. Ryzhik, Tables of Integrals, Series and Products, Academic Press, Sydney, Australia, 1980.

[14] L.G. Sillen, A.E. Martell, Stability Constants of Metal-ion Complexes, Chemical Society, London, UK, 1964.

[15] R. Pruiksma, R.L. McCreery, Observation of electrochemical concentration profiles by absorption spectroelectrochemistry, Anal.Chem. 51(1979) 2253-2257.

[16] C.C. Jan, R.L. McCreery, High-resolution spatially resolved visible absorption spectrometry of the electrochemical diffusion layer, Anal.Chem. 58(1986)2771-2777.

[17] G. Chen, R.L. Fournier, S. Varanasi, A mathematical model for the generation and control of a pH gradient in an immobilized enzyme system involving acid generation, Biotechnology and Bioengineering. 57(1998) 394-407.

**Figure Captions**

Figure 1 (a)

[$Fe^{2+}$] versus time. Note that it rises from zero sharply and becomes steady within a short time. [Note that the concentration unit used throughout the figures is mole/dm$^3$]

Figure 1(b)

[$Fe(OH)^+$] versus time. The rise is gradual.

Figure 1 (c)

[Fe(OH)$_2$] versus time.

Figure 1 (d)

[H$^+$] versus time. Note the sudden fall.

Figure 1 (e)

[OH$^-$] versus time. The rise is gradual.

Figure 1(f)

The ionic product of water [H$^+$] x [OH$^-$] versus time. It is independent of time as it should.

Figure 1(g)

[Fe$^{2+}$]x[OH$^-$]$^2$x10$^{17}$ versus time. The horizontal line is the solubility product of ferrous hydroxide. The intersection gives the passivation time which is nearly 6 hours.

Figure 1(h)

[H$_2$O] versus time. Note that it changes only in the 4-th decimal place as expected.

Figure 2

Plots of the concentration product [Fe$^{2+}$]x[OH$^-$]$^2$x10$^{17}$ versus time for the free corrosion condition with different corrosion currents: (a) 2.08 * 10^ (-5) A/ dm$^2$ (b) 2.08 * 10^ (-4) A/ dm$^2$ (c) 2.08 * 10^ (-3) A/ dm$^2$. The solid horizontal line is the solubility product of ferrous hydroxide.

Figure 3

Plots of the concentration product [Fe$^{2+}$]x[OH$^-$]$^2$x10$^{17}$ versus time for cathodic polarization, free corrosion and anodic polarization conditions with different impressed currents[IC]: - 1.04 * 10 ^ (-4) A/dm$^2$ ;

-0.52* 10 ^ (-4) A/dm$^2$; Free Corrosion ; + 0.52* 10 ^ (-4) A/dm$^2$ ; +1.04 * 10 ^ (-4) A/dm$^2$ ;   the solid horizontal line is the solubility product of ferrous hydroxide. The point of intersection with the horizontal line gives the passivation time.

Figure 4

Concentration versus distance from the corroding metal surface at time=6.9 hrs which is slightly more than the passivation time: (a) [Fe$^{2+}$]  (b) a slight increase in the water concentration (c) [Fe(OH)$^+$]  (d) [Fe(OH)$_2$]  (e) [Fe$^{2+}$]x[OH$^-$]$^2$x10$^{17}$  (f)  [Fe$^{2+}$]x[OH$^-$]$^2$x10$^{17}$ at time=6.1 hours which is slightly less than the passivation time.

Figure 5

The zinc system: plots of the concentration product [Zn$^{2+}$]x[OH$^-$]$^2$x10$^{17}$ versus time for the free corrosion condition with different corrosion currents: (a) 4.58 * 10^ (-3) A/ dm$^2$  (b) 4.58 * 10^ (-4) A/ dm$^2$  (c) 4.58 * 10^ (-5) A/ dm$^2$ . The solid horizontal line is the solubility product of zinc hydroxide.

**Figures**

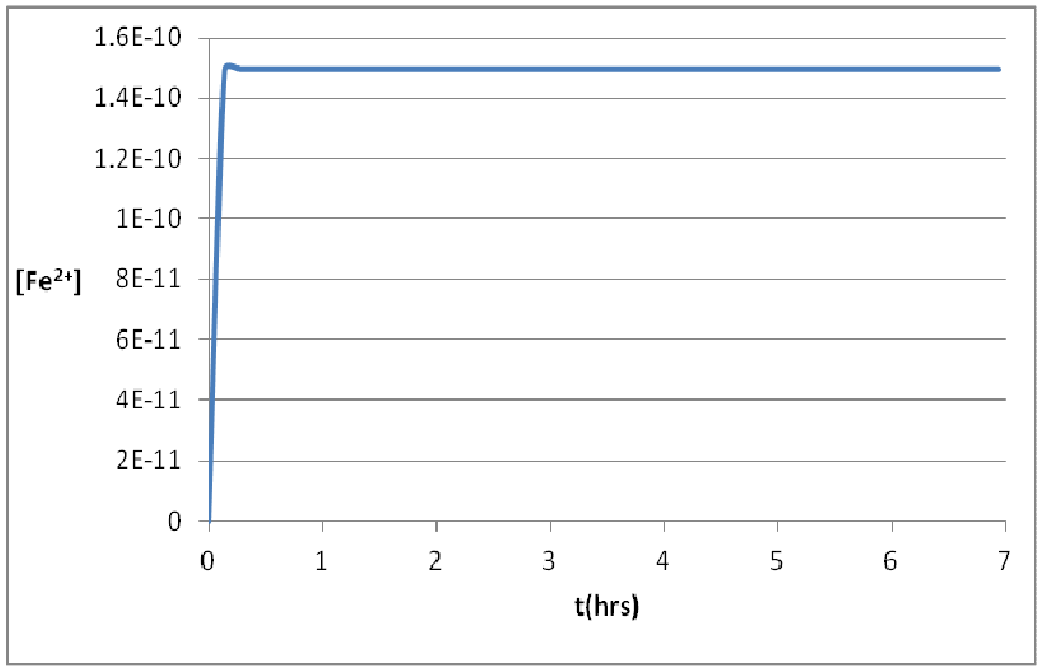

Figure-1(a)

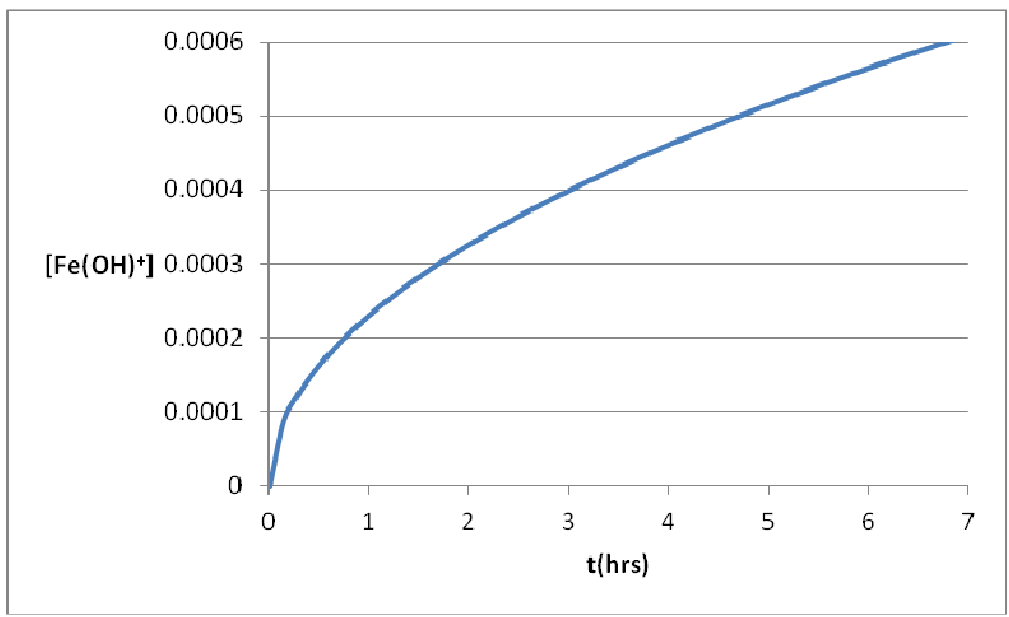

Figure-1(b)

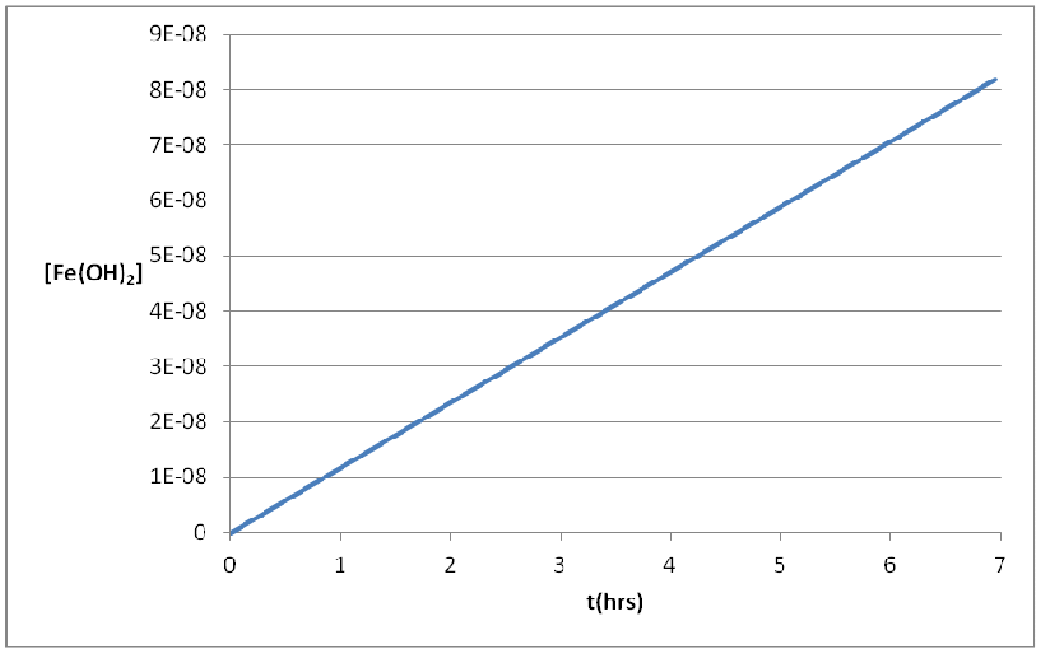

Figure-1(c)

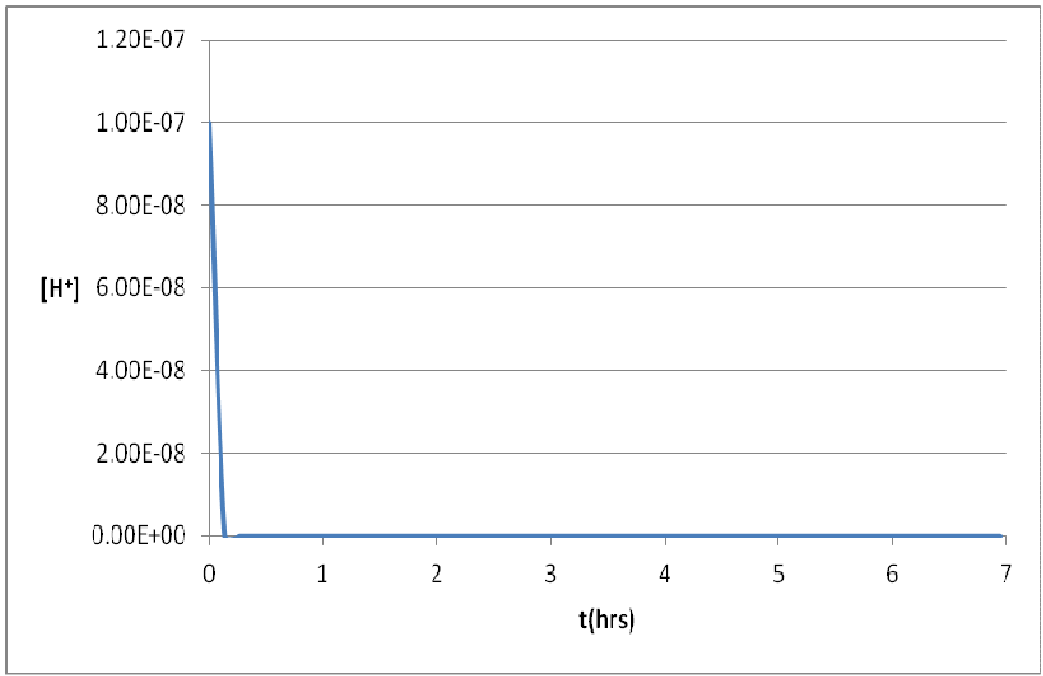

Figure-1(d)

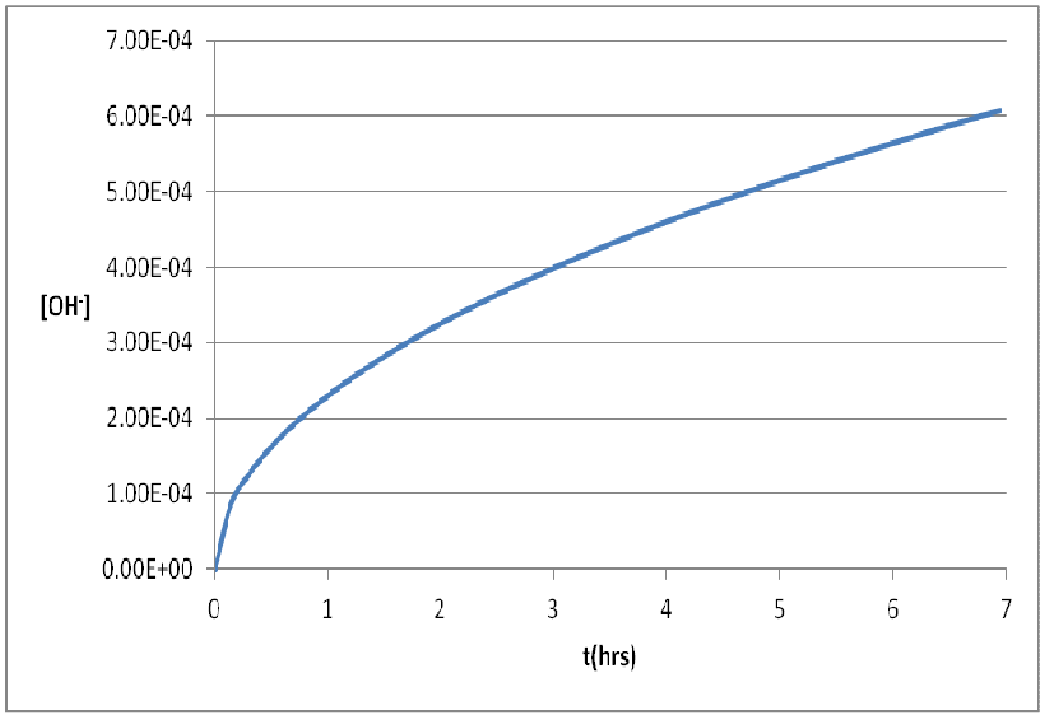

Figure-1(e)

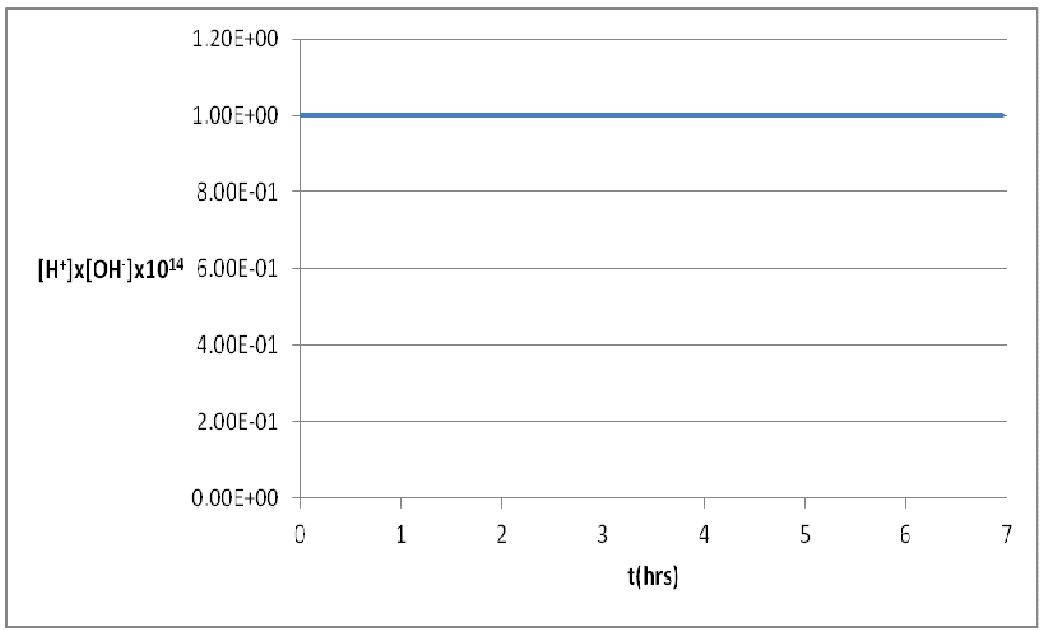

Figure-1(f)

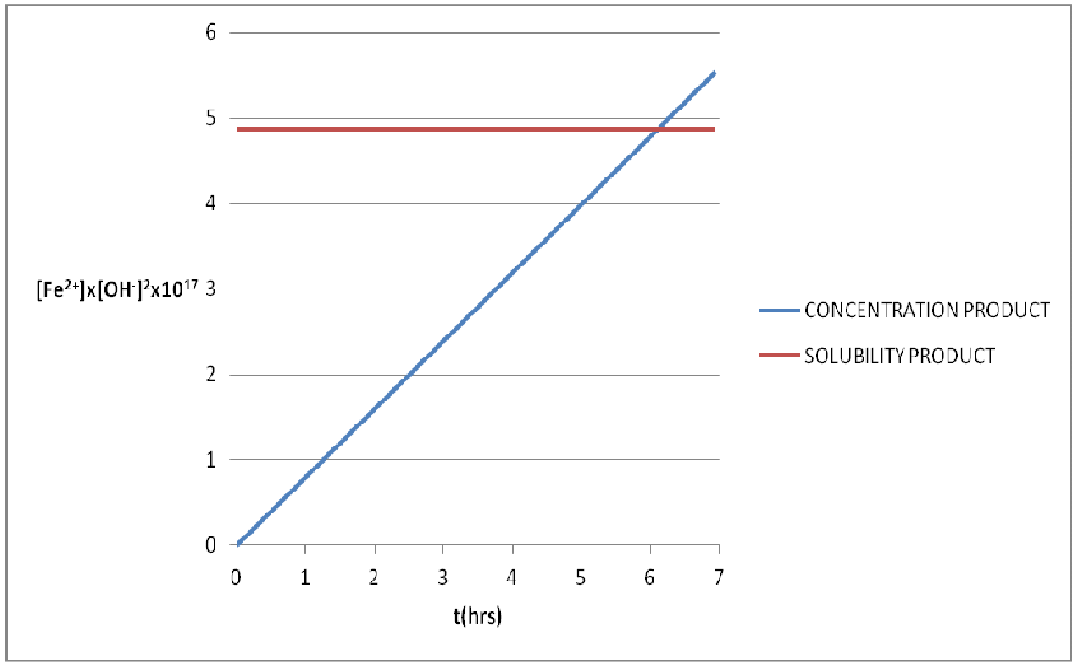

Figure-1(g)

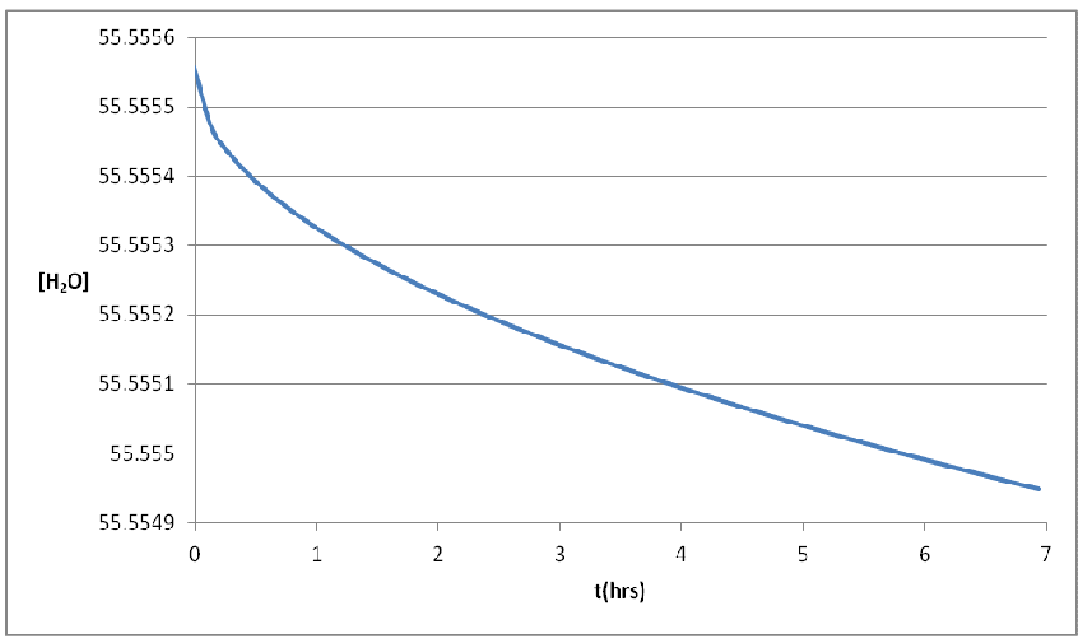

Figure-1(h)

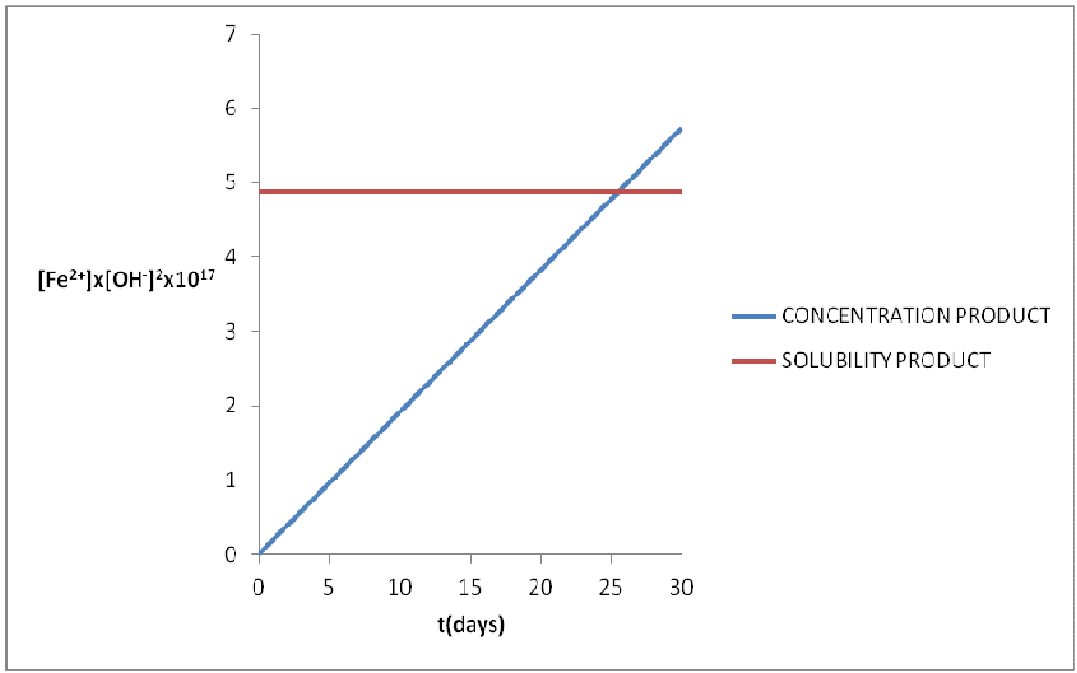

Figure-2(a)

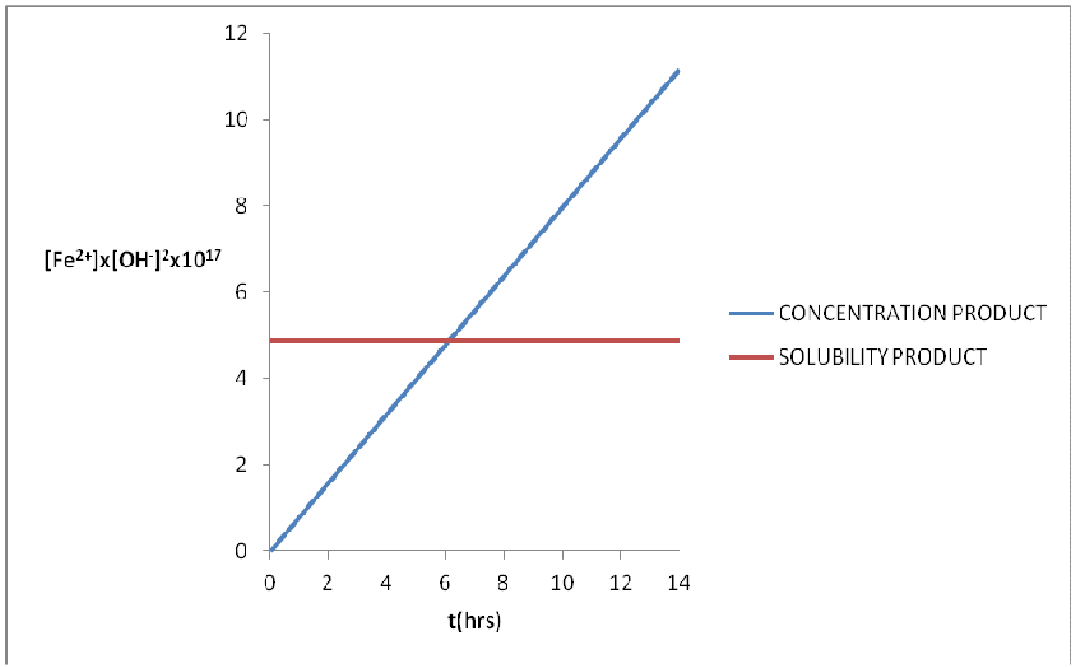

Figure-2(b)

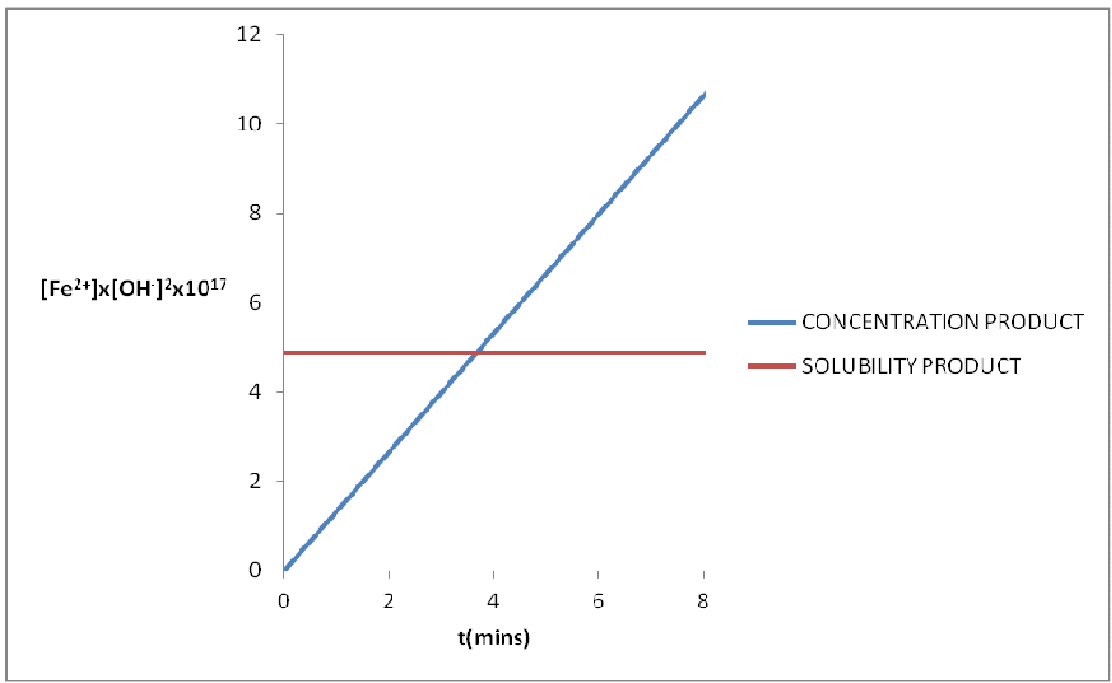

Figure-2(c)

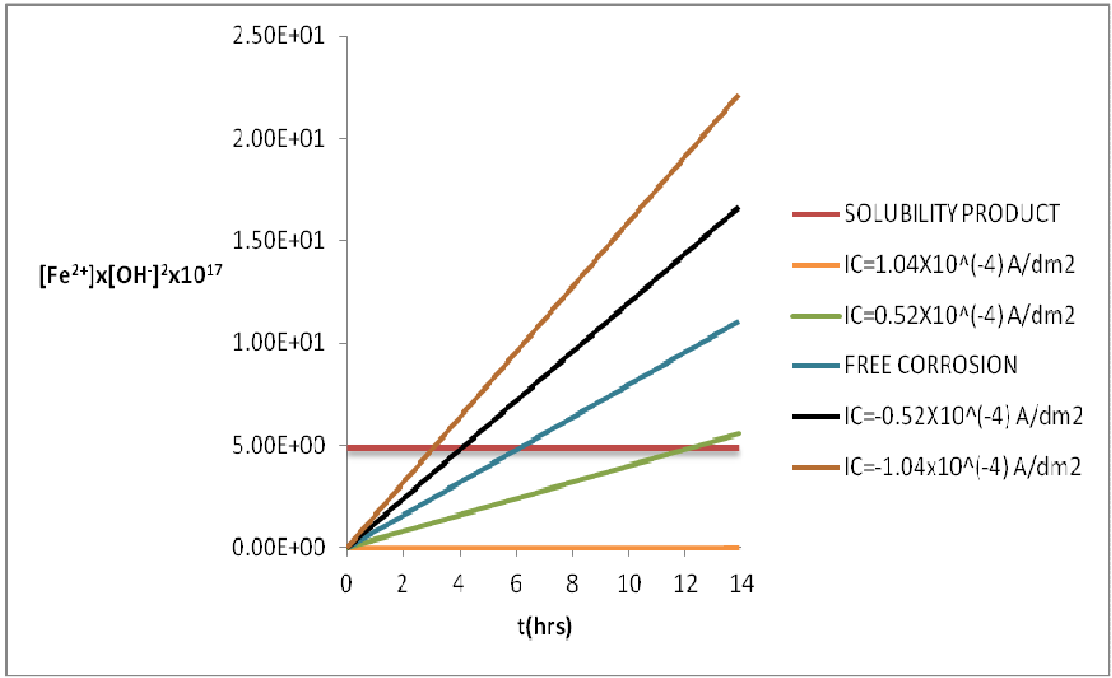

Figure-3

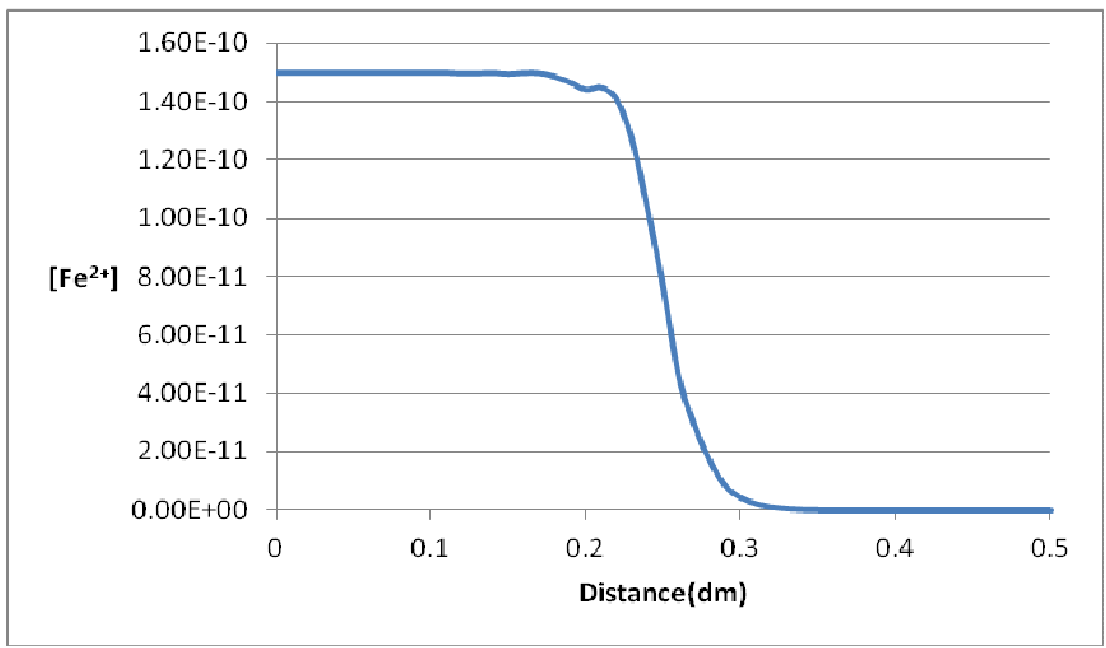

Figure-4(a)

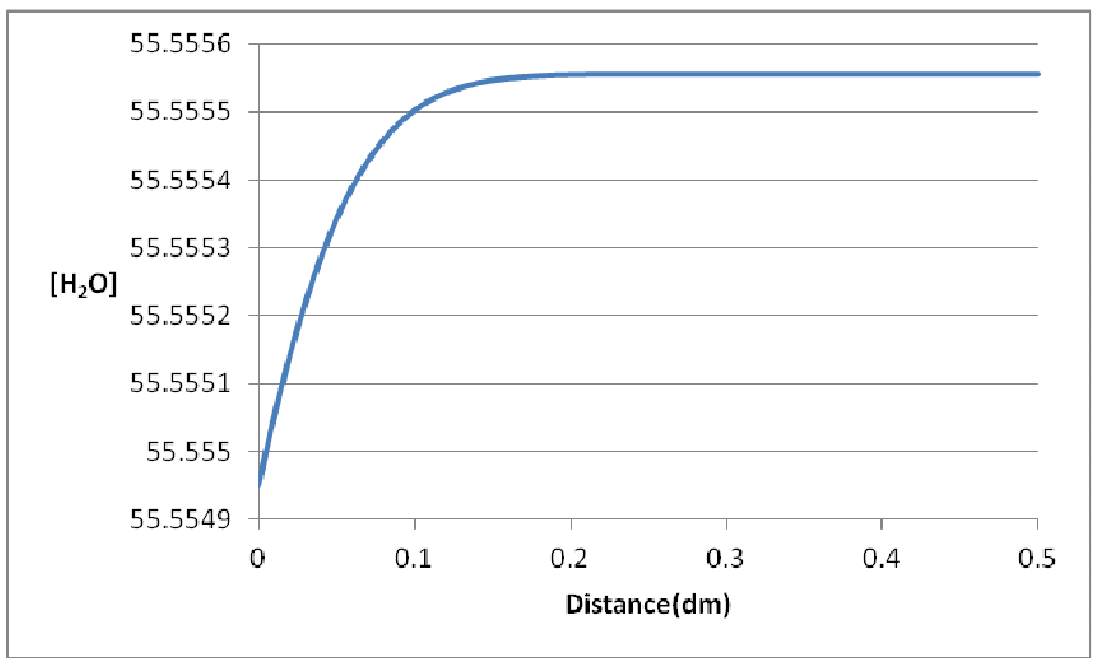

Figure-4(b)

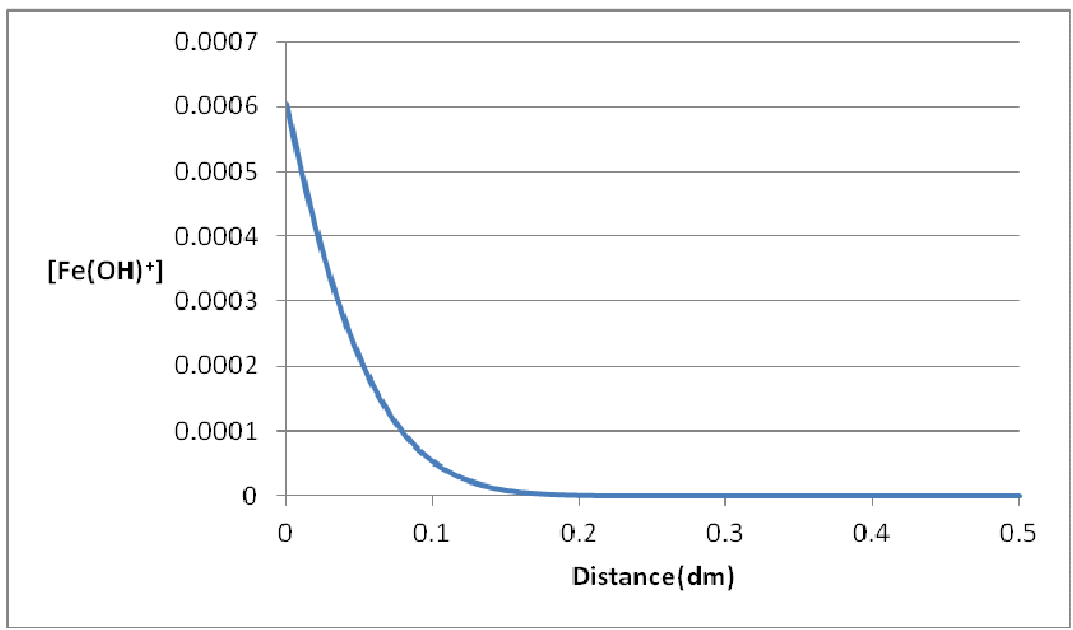

Figure-4(c)

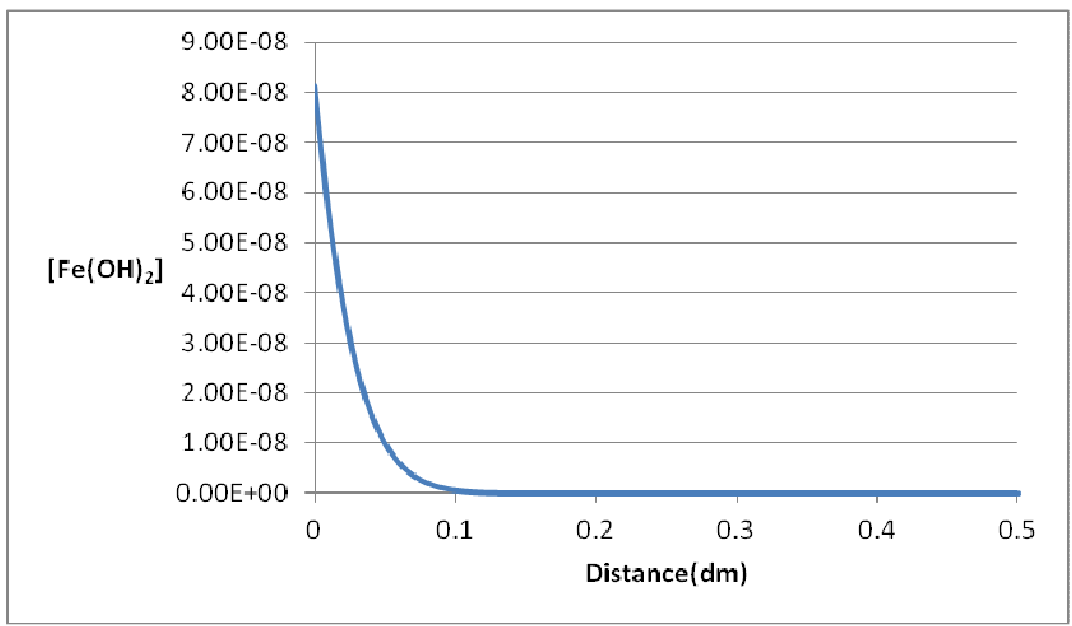

Figure-4(d)

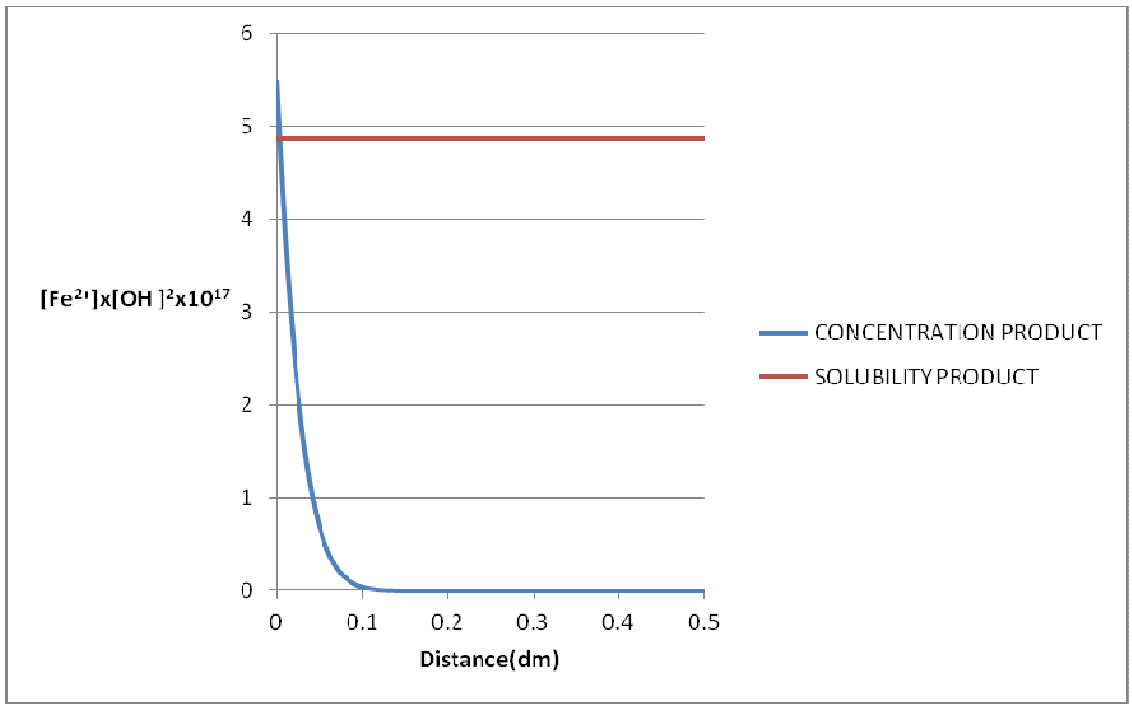

Figure-4(e)

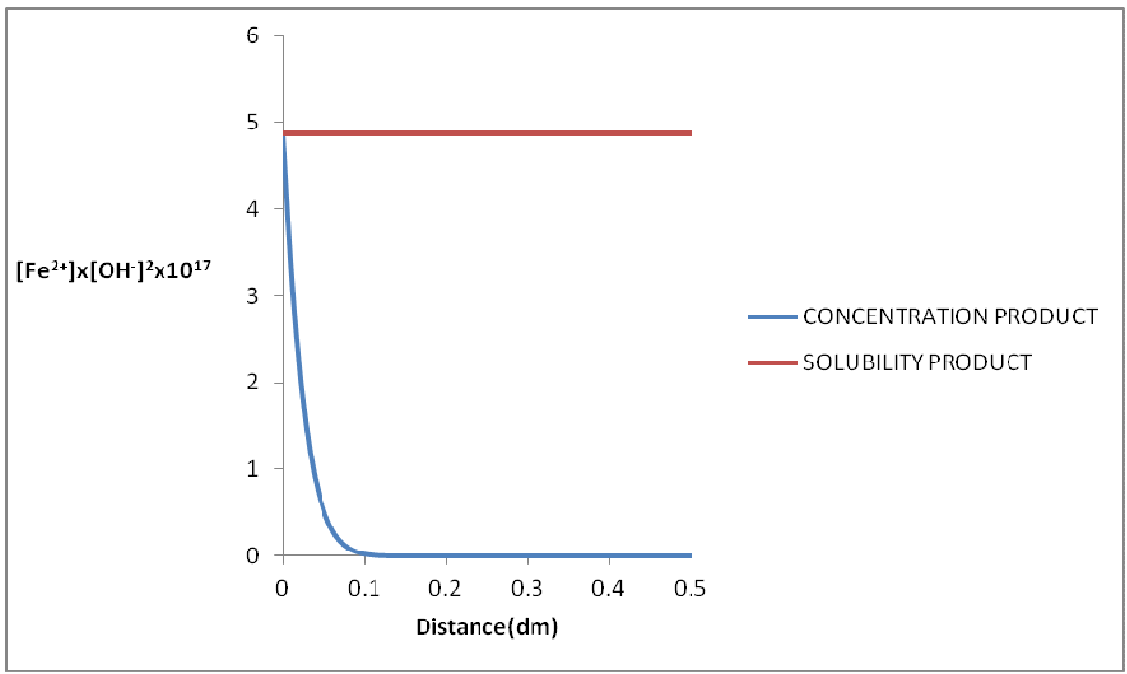

Figure-4(f)

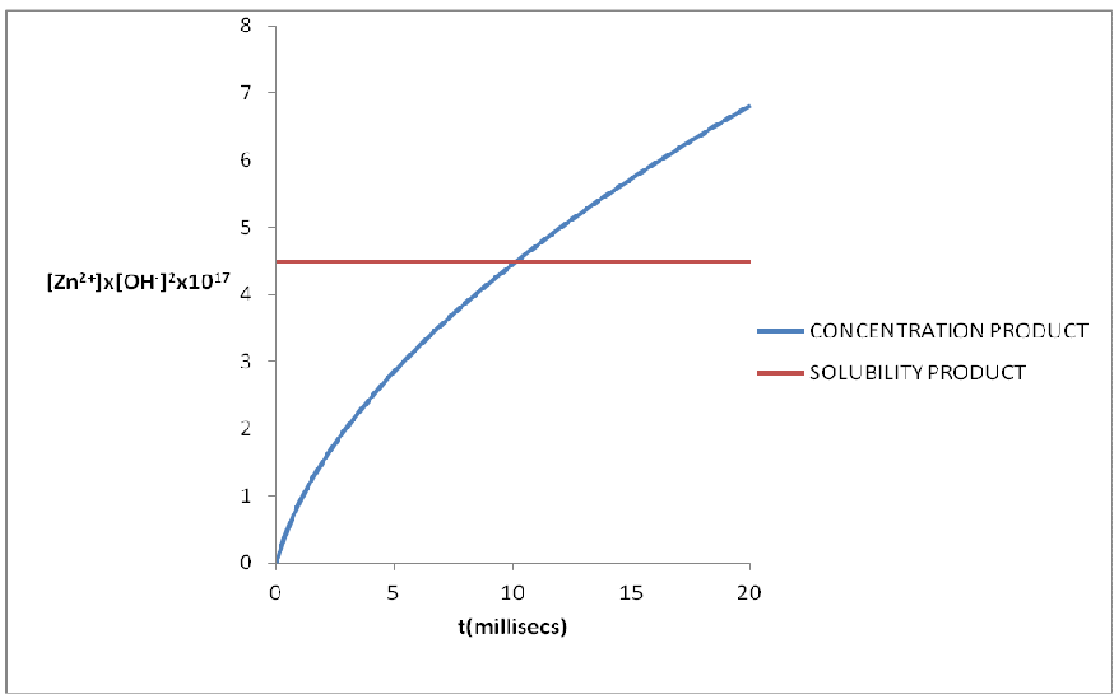

Figure-5(a)

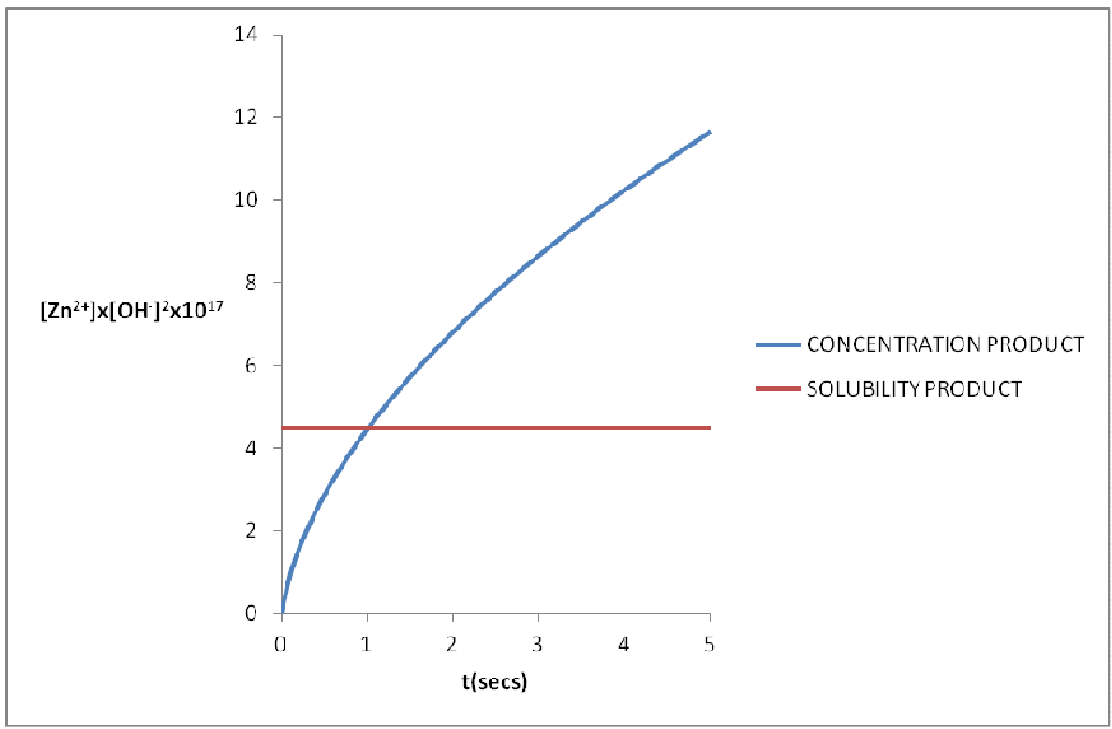

Figure-5(b)

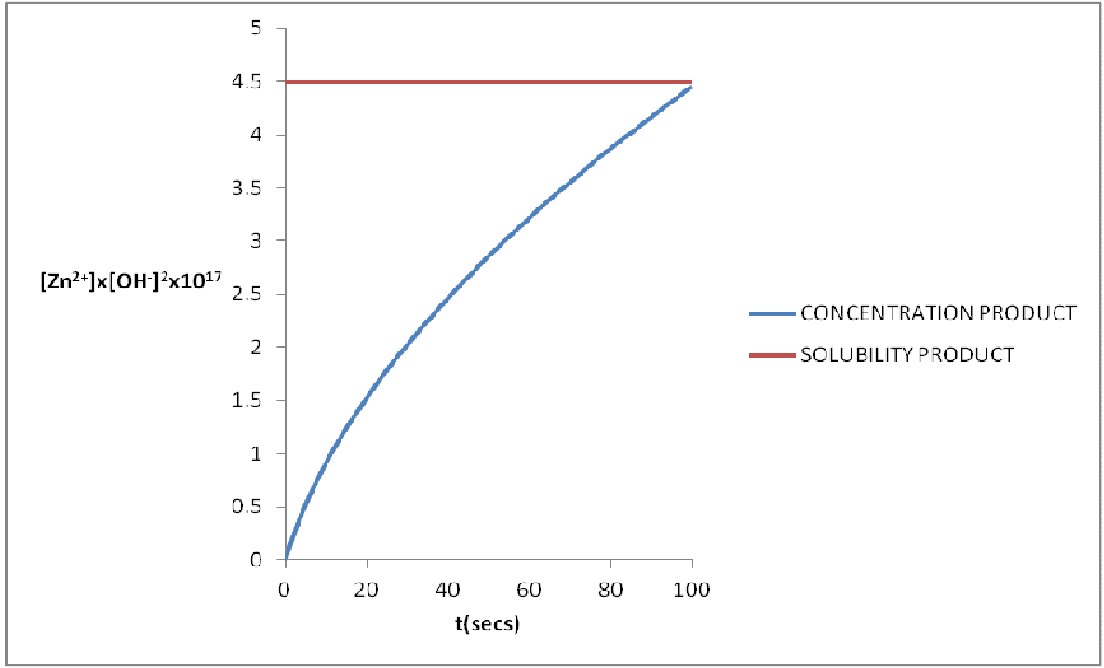

Figure-5(c)

**Graphic Image**

---

$M \rightarrow M^{n+} + ne \qquad O_2 + 2H_2O + 4e \rightarrow 4OH^-$

$M^{2+} + H_2O \Leftrightarrow (MOH)^+ + H^+ \qquad \dfrac{\partial C_1}{\partial t} = D_1 \dfrac{\partial^2 C_1}{\partial x^2} - R_1$

$(MOH)^+ + H_2O \Leftrightarrow M(OH)_2 + H^+ \qquad \dfrac{\partial C_2}{\partial t} = D_2 \dfrac{\partial^2 C_2}{\partial x^2} - (R_1 + R_2 + R_3)$

$H_2O \Leftrightarrow H^+ + OH^- \qquad \dfrac{\partial C_6}{\partial t} = D_6 \dfrac{\partial^2 C_6}{\partial x^2} + R_3$

---